\documentclass{optica-article}

\journal{opticajournal} 

\articletype{Research Article}
\usepackage{graphicx,epsf,amsmath}
\usepackage{dcolumn}
\usepackage{bm,bbm}
\usepackage{enumerate}
\usepackage[normalem]{ulem}
\usepackage{upgreek}
\usepackage{braket}
\usepackage{siunitx}
\newcommand{\op}[2]{|#1\rangle\langle#2|}

\usepackage{float}
\usepackage{dsfont}
\usepackage{mathtools}

\usepackage{graphicx}
\usepackage{hyperref}
\usepackage{cleveref}
\usepackage{subcaption} 
\usepackage{lineno}

\begin{document}
\title{Feasibility study of frequency-encoded photonic qubits over a free-space channel}
\author{St\'ephane Vinet,\authormark{1,*} Wilson Wu,\authormark{1,2} Yujie Zhang,\authormark{1} and Thomas Jennewein\authormark{1,2}}

\address{\authormark{1}Institute for Quantum Computing and Department of Physics \& Astronomy,
University of Waterloo, 200 University Ave W, Waterloo, Ontario, N2L 3G1, Canada\\
\authormark{2}Department of Physics, Simon Fraser University, 8888 University Dr W, Burnaby, BC V5A 1S6, Canada}
\email{\authormark{*}svinet@uwaterloo.ca}

\begin{abstract*} 
Frequency-bin quantum encoding shows great promise for quantum communication given its high-dimensional scaling, compatibility with photonic integrated circuits and synergy with classical optical communication technology. However, to date all demonstrations have been performed over single-mode and static channels, while the transmission over fluctuating and turbulent channels has not been addressed. 
We propose and demonstrate a novel approach that leverages field-widened interferometers to decode frequency-bins transmitted over free-space channels without any adaptive optics or modal filtering. Moreover, we investigate the phase stability requirements so that frequency-bin encoding could be feasible for satellite to ground quantum links. Our passive approach expands the versatility of frequency-bin encoding, paving the way towards long-range and fluctuating  channels. 
\end{abstract*}
\section{Introduction}
Quantum communication has garnered significant interest due to its ability to provide secure information transmission relying on the principles of quantum mechanics \cite{RevModPhys.81.1301}. Frequency-bin encoding is a promising approach for quantum information encoding due to its ease of implementation, compatibility with integrated photonic devices\cite{Imany:18,PhysRevA.82.013804,PhysRevA.88.032322,Lukens:17}, and its high dimensionality which allows for large alphabets \cite{clementi_programmable_2023, Cabrero,PhysRevApplied.19.064026,kues_-chip_2017}.  This is particularly beneficial for quantum communication as it increases the capacity of the quantum channel and enhances noise tolerance \cite{PhysRevX.9.041042,Dixon2012,cozzolino2019}. Frequency-bin states can be generated from compact sources, are easily manipulated using standard optical components, and the stability in their relative phase provides inherent resistance against mode mixing and phase noise in long-distance propagation. Furthermore, frequency-bin encoding has shown promise in overcoming challenges associated with solid state emitters, such as phonon dephasing and spectral diffusion, thus allowing for the generation of photonic cluster states \cite{PhysRevA.98.022318,pennacchietti_oscillating_2024} and facilitating scalable quantum information processing \cite{PhysRevA.105.062445,laccotripes_spin-photon_2024,Lu:23,gao_quantum_2013}. In particular, frequency-bins are compatible with spectrally heterogeneous matter qubits \cite{Lingaraju:22,Lukens:17} and support dense spectral multiplexing making them prime candidates for quantum networking.
Frequency-bin states are typically decoded via spectral filtering \cite{PhysRevA.82.013804,Bloch:07,tagliavacche2024frequencybinentanglementbasedquantumkey}, electro-optic modulation \cite{Imany:18,Merolla:99,Olislager_2012}, or pulse shaping \cite{PhysRevA.88.032322,lu_bayesian_2022,Cabrero}. However, these techniques often introduce losses or require single-mode coupling. In this paper, we propose a simple interferometric decoding method utilizing a field-widened Mach-Zehnder interferometer (MZI) which requires no modal filtering, adaptive optics or active elements, thus providing a robust solution for multimode channels and free-space quantum communication \cite{PhysRevA.97.043847}. This passive interferometric approach offers important advantages over the established techniques for manipulating frequency-bin encoded quantum states by leveraging the inherent phase relationships between frequency components. In contrast with spectral filtering, which sequentially isolates separate frequencies, interferometric methods mix frequency components through interference, enabling simultaneous decoding of multiple frequency-bins.
\section{Theory}\label{Theory}
We first consider an unbalanced Mach-Zehner interferometer, with a path length difference $\Delta L$ chosen so  it  acts as a frequency demultiplexer (DEMUX) that directs specific frequencies to designated output ports based on their relative phases according to Eq.~\ref{eq:MZI}:
\begin{equation}\label{eq:MZI}
   \Delta \omega=\frac{\pi c}{\Delta L},
\end{equation}
where $\Delta \omega$ is the DEMUX frequency separation.
This method extends naturally to qudits by concatenating a series of optical interleavers \cite{PhysRevLett.130.200602, erhard_advances_2020} or using cascaded interferometers \cite{PhysRevApplied.7.044010,Brougham_2013}. However, because the spatial and temporal modes of a photon are distorted after travelling through a multi-mode optical channel\cite{Dikmelik:05}, these distortions introduce path distinguishability thereby impeding the interference visibility. This is exacerbated for mobile links as telescope pointing errors and turbulence induced angular fluctuations further degrading the interference quality. Moreover, spatial filters cannot address these issues without introducing significant throughput losses.
 Jin \textit{et al.} \cite{PhysRevA.97.043847,Jin:19,tannous2023fullypassivetimebinquantum,10.1117/12.2218282,7425090,PhysRevLett.116.253601} demonstrated that field-widened interferometers can effectively be used as multi-mode time-bin quantum receivers by exploiting imaging optics, or carefully selected refractive indices, to correct induced phase shifts and visibility reductions due to angular fluctuations, which we realize can be extended to the analysis of frequency-bin receivers. Using relay optics in both arms of the unbalanced Mach-Zehnder interferometer, the differences in the evolution of the spatial modes over the path length difference $\Delta L$ can be matched thus guaranteeing identical wavefronts when the paths recombine, ensuring indistinguishability at the interferometer output regardless of the spatial mode or the angle of incidence of the input beam. 
\par
\begin{figure}[ht]
    \centering
        \begin{subfigure}{.499\textwidth}
      \centering
      \includegraphics[width=0.9\textwidth]{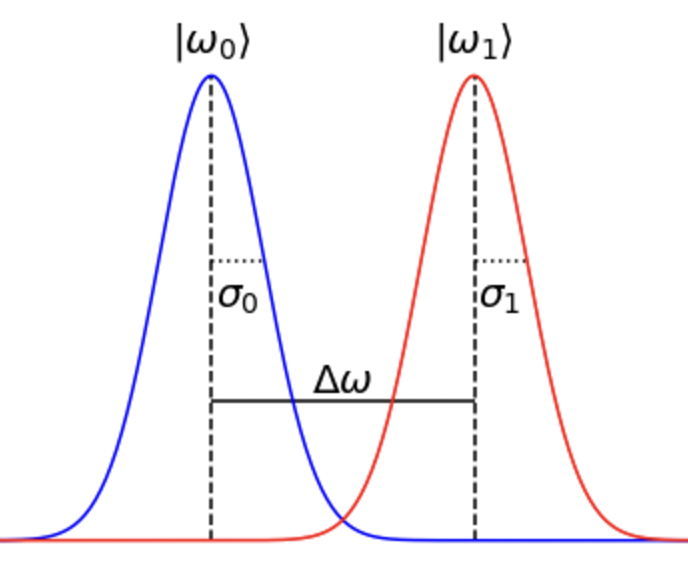}
      
      \caption{}
         \label{fig:encoding}
    \end{subfigure}%
        \begin{subfigure}{.499\textwidth}
      \centering
       \includegraphics[width=0.98\textwidth]{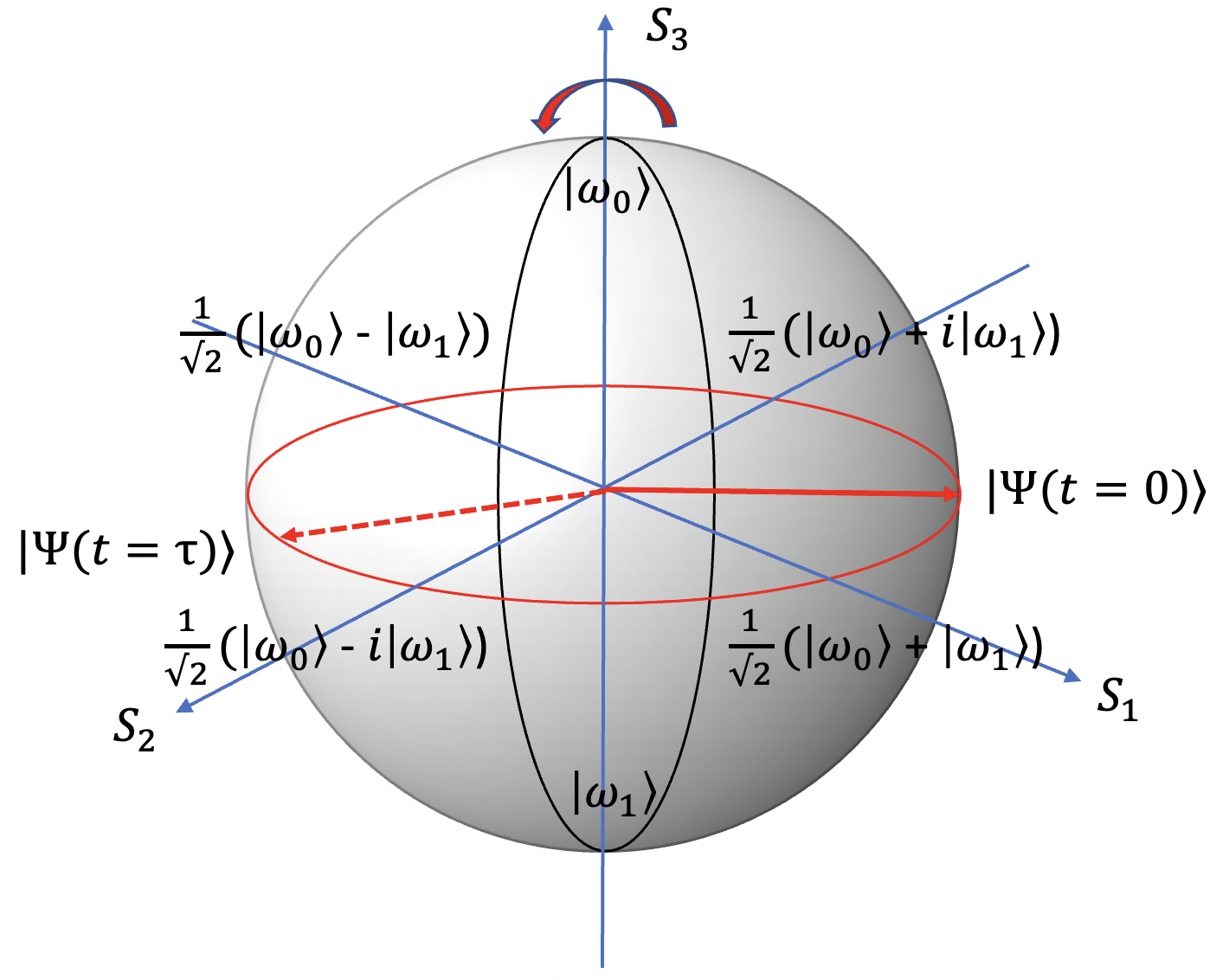}
      \caption{}
                \label{fig:poincare}
          \end{subfigure}
          \caption{(a)  Frequency-bin encoding. Photons are encoded in spectral bands $\ket{\omega_0}$ ($\ket{\omega_1}$) with bandwidth $\sigma_0$ ($\sigma_1$) separated by $\Delta\omega$. For $\braket{\omega_0|\omega_1}=0$, $\ket{\omega_0}$ ($\ket{\omega_1}$) correspond to logical bits $\ket{0}_L$ ($\ket{1}_L$). (b) Poincaré sphere. The time dependence, described in Schrödinger picture, $\ket{\Psi(t)}=\frac{1}{\sqrt{2}}(\ket{\omega_0}+e^{i\Delta\omega t}\ket{\omega_1})$ leads to an oscillation around the equator at angular velocity $\Delta\omega=\omega_1-\omega_0$ as depicted at times $t=0$ and $t=\tau$.}
   \label{}
\end{figure}

 In our scheme, we consider frequency-encoded photonic qubits, defined as: $\ket{\omega_0}$, $\ket{\omega_1}\coloneqq\ket{\omega_0+\Delta\omega}$, and the superposition $\frac{1}{\sqrt{2}}[\ket{\omega_0}+e^{i\Delta\omega t}\ket{\omega_1}]$, corresponding to logical bits $\ket{0},\ket{1}, \ket{+}$ for $e^{i\Delta\omega t}=1$. For the superposition (\(X\)) basis, different frequency-bins accumulate phases at a different rate leading to a time dependent relative phase. Consequently, in the Schrödinger picture the time evolution leads to an oscillating beat note $\ket{\Psi(t)}=\frac{1}{\sqrt{2}}[\ket{\omega_0}+e^{i\Delta\omega t}\ket{\omega_1}]$ around the equator of the Poincaré sphere as depicted in Fig~\ref{fig:poincare} \cite{PhysRevLett.124.190502}. 
In spite of the phase oscillation of the quantum state, Pennacchietti et al. have shown that when the temporal resolution of the detection system is significantly smaller than the oscillation period, all photons can be used in a quantum key distribution protocol for key generation \cite{pennacchietti_oscillating_2024}. 
\subsection{Monochromatic frequency-bins}
For a superposition $\ket{\Psi}$ of two frequency-bins, $(A_0\hat{a}_0^\dagger+A_1\hat{a}_1^\dagger)\ket{0}$, the intensity measurement $\langle \hat{N}(t)\rangle$, where $\hat{N}(t)=\hat{a}_t^\dagger \hat{a}_t$, results in a beat note \cite{feynman1963}
\[\langle \hat{N}(t)\rangle=\frac{1}{(|A_0|+|A_1|)^2}\bigg(|A_0|^2+|A_1|^2+2|A_0||A_1|\cos(\Delta\omega t)\bigg).\]The corresponding beat visibility is $v_X=\frac{2 |A_0 A_1|}{|A_0|^2 + |A_1|^2 }\cdot v_{X}^\epsilon,$ where $v_{X}^\epsilon$ corresponds to the visibility due to mode-mismatch and experimental imperfections for the \(X\)-basis measurement. \par For the rectilinear (\(Z\)) basis, the phase shift introduced by the unbalanced MZI is given by the product of the spatial frequency and the path length difference $\phi_i=\omega_i\Delta L/c$ leading to a phase difference at the output equal to $\Delta\phi=\phi_1-\phi_0=\frac{\Delta\omega\Delta L}{c}=\pi$ when $\Delta L=\pi c/\Delta \omega$ in accordance with Eq.~\ref{eq:MZI}.
\subsection{Non-monochromatic frequency-bins}
In practice, each frequency mode $\ket{\omega_i}$ is not perfectly monochromatic but has a finite bandwidth $\sigma_i$. Assuming Gaussian wave packets, each mode can be described as
\begin{equation}
\ket{\omega_i}=\bigg(\frac{1}{2\pi\sigma^2_i}\bigg)^{1/4}\int d\mu \exp\bigg(-\frac{(\mu-\omega_i)^2}{4\sigma_i^2}\bigg)\hat{a}^{\dagger}_{\mu}\ket{0}
\end{equation}
with the creation operator $\hat{a}^{\dagger}_{\mu}$ for a monochromatic frequency mode $\mu$ as shown in Fig.~\ref{fig:encoding}. Consequently, a realistic frequency-bin superposition state can be described as
\begin{align}\label{eq:chromatic}
\ket{\Psi}&=\int d\mu \left[A_0\Big(\frac{1}{2\pi\sigma^2_0}\Big)^{1/4}\exp\Big(-\frac{(\mu-\omega_0)^2}{4\sigma^2_0}\Big) 
 +A_1\Big(\frac{1}{2\pi\sigma^2_1}\Big)^{1/4}\exp\Big(-\frac{(\mu-\omega_1)^2}{4\sigma^2_1}\Big)\right]\hat{a}^{\dagger}_{\mu}\ket{0}\\
 &=A_0\ket{\omega_0}+A_1\ket{\omega_1}
 \approx A_0\ket{0}_L+A_1\ket{1}_L \nonumber,
\end{align}
where $|A_0|^2+|A_1|^2=1$  and the approximation $\ket{\Psi}\approx A_0\ket{0}_L+A_1\ket{1}_L$ holds when $\braket{\omega_0|\omega_1}\simeq 0$. The inner product
\begin{equation*}
    \braket{\omega_0|\omega_1}=\sqrt{\frac{2\sigma_0\sigma_1}{\sigma^2_0+\sigma^2_1}}\exp\bigg(-\frac{\Delta\omega^2}{4(\sigma^2_0+\sigma^2_1)}\bigg)
\end{equation*}
has minimal overlap when the frequency separation $\Delta\omega=\omega_1-\omega_0$ is much larger than the frequency-bin bandwidth (i.e. $\Delta\omega\gg2\sqrt{\sigma^2_0+\sigma^2_1}$).
We now consider an intensity measurement $\hat{N}(t)=\hat{a}_{t}^{\dagger}\hat{a}_{t}$ with $\hat{a}_t=\int d\mu \hat{a}_{\mu}e^{i\mu t} $ on the state in Eq.~\ref{eq:chromatic} and obtain a time-dependent beating signal
\begin{align}
F(t)\propto \bra{\Psi}\hat{N}\ket{\Psi}:&= \frac{1}{2}(1+v_X'\cos(\Delta\omega t)),
\end{align}
where $v_X'=v_X(\sqrt{2\sigma_0\sigma_1}e^{-t^2(\sigma_0^2+\sigma_1^2)})/(\sigma_0e^{-2t^2\sigma_0^2}+\sigma_1e^{-2t^2\sigma_1^2})$ . 
\par In the \(Z\)-basis, the phase shift introduced by the path length difference in the MZI depends on the bandwidth of the frequency-bin according to $\delta\phi_i=2\pi\sigma_i\Delta L/c$ \cite{Born_Wolf_2019}. Assuming the phase shift in each frequency-bin follows a Gaussian distribution, the total phase shift variation across the two frequency-bins is then equal to $\delta\phi_T=\sqrt{\delta\phi^2_0+\delta\phi^2_1}$ and its impact on the MZI visibility corresponds to 
\begin{equation*}
    v_Z= v_{Z}^{\epsilon}\exp\bigg({\frac{-2\pi^4(\sigma^2_{0}+\sigma^2_{1})}{\Delta\omega^2}}\bigg),
\end{equation*}
where $v_{Z}^\epsilon$ is the visibility due to the experimental imperfections for the \(Z\)-basis measurement. For practical implementations, $\sigma_0 \approx \sigma_1$ and the source's linewidth $\sigma_i\ll\Delta\omega$ is several orders of magnitude smaller than the frequency-bin separation. These bandwidth contributions are thus negligible.

\subsection{Timing jitter}
\begin{figure}[ht]
    \centering
        \begin{subfigure}{.499\textwidth}
      \centering
      \includegraphics[width=\textwidth]{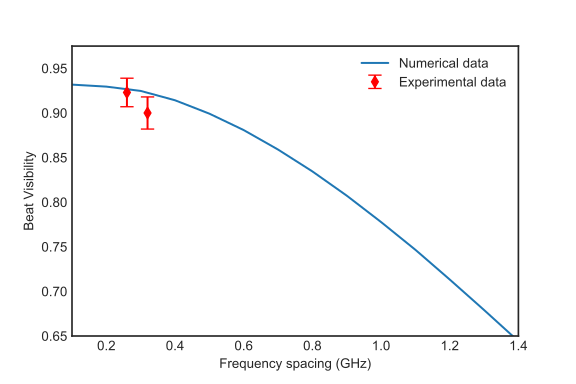}
      \caption{Frequency spacing limitation}
         \label{fig:spacing}
    \end{subfigure}%
        \begin{subfigure}{.499\textwidth}
      \centering
       \includegraphics[width=\textwidth]{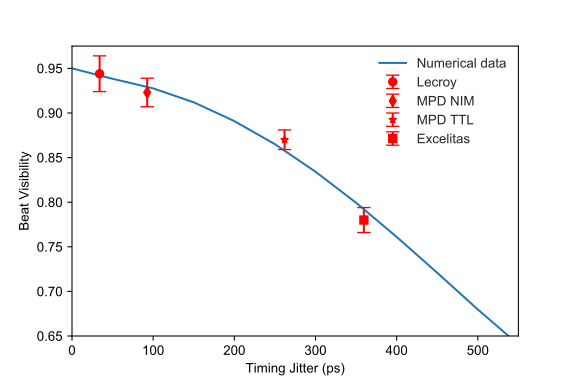}
      \caption{Timing resolution limitation}
         \label{fig:visi}
          \end{subfigure}
\caption{Limitations on the beat note visibility due to (a) the beat note's frequency spacing for a fixed timing resolution of 100 ps, (b) the detector's time jitter for a fixed frequency spacing of $260$ MHz. We assume a laser linewidth of 300 kHz \cite{Toptica2024}, and a visibility of $95\%$ at $\Delta t=0$.  }
   \label{fig:jitter}
\end{figure}

The phase oscillation of the superposition state shown in Fig.~\ref{fig:poincare} constrains the resolvable frequency spacing $\Delta \omega$ due to detector timing jitter. Fig.~\ref{fig:spacing} shows the maximal visibility for a timing resolution of 100 ps as a function of $\Delta\omega$ assuming $v_X=0.95$. Alternatively for a given frequency spacing, this constraint reduces the beat note visibility as a function of timing jitter as shown in Fig.~\ref{fig:visi}.  To model timing jitter, the measurement outcome
$F(t)$ can be described as a convolution of the beating signal $f(t)=\frac{1}{2}(1+v_X\cos(\Delta\omega t))$  with weight function $g(t)$, where $v_X$ is the initial visibility of the state. Assuming Gaussian $g(t)$ with timing jitter standard deviation of $\delta t_j$ and adding sources of jitter in quadrature ($\delta T=\sqrt{\sum_j\delta t_j^2}$) the measurement outcome $F(t)$ is
 \begin{equation}
     F(t)=\int g(\tau)f(t-\tau)d\tau =\int f(t-\tau)\frac{1}{2\sqrt{2\pi}}\exp[-\frac{\tau^2}{2\delta T^2}]d\tau=\frac{1}{2}[1+v_Xe^{-(\Delta\omega\delta T)^2/2}\cos(\Delta\omega t)].
 \end{equation}
The visibility of the beating signal is degraded as $v^{\rm jitter}_X= v_X\exp[{-(\Delta\omega\delta T)^2/2}]$. 
 We experimentally confirm this result by measuring a $260$ MHz beat note using the experimental setup in Fig.~\ref{fig:setup} with four single photon detectors each with a different timing resolution given in Table~\ref{tab:jitter}. As the photon count rate is significantly smaller than the number of beat cycles $N_b$, we observe the beat note via statistical accumulation by post-processing the time-tags, referencing each detection event $(t_i)$ to the latest synchronization marker pulse $(t_m)$ according to 
\begin{equation}\Delta t_i=t_i-\max_m \{t_m|t_m\leq t_i\},\end{equation}
where the synchronization channel provides a marker pulse with frequency $1/T_M$. Each time difference $\Delta t_i$ is mapped into the beat period $T_b=T_M/N_b=2\pi/\Delta \omega$ via 
\begin{equation}\tau_i=\Delta t_i\mod{T_b},\end{equation}
effectively assigning a time relative to the beat period to each photon detection event. Partitioning the beat period into $K=T_b/\delta t_{\rm bin}$ time intervals, we calculate the number of events falling within the $k$-th interval according to 
\begin{equation}N_k(T)=\sum_i\mathbf{1}_A(\tau_i),\end{equation}
where $\mathbf{1}_A$ is the indicator function and $A=[k\delta t_{\rm bin},(k+1)\delta t_{\rm bin}]$ represents the $k$-th bin. Summing over all intervals $\delta t_{\rm bin}$, we create a histogram of the total events per beat period 
\begin{equation}N(T)=\sum_{k=0}^{K-1}N_k(T)\end{equation}
where the bin size $\delta t_{\rm bin}$ is the resolution of the time-tagger. 
\begin{table}
    \centering
   \captionsetup{justification=centering}
        \caption{Detector models used in Fig.~\ref{fig:visi} and their corresponding nominal FWHM timing resolution ($\delta t^{\rm FWHM}$). The MPD and Excelitas detectors were used in pair with a UQD Logic16 time-tagger with a resolution of $78.125$ ps \cite{UQDevices_Logic16_Datasheet}. A conservative upper bound for the timing resolution of the LeCroy waveRunner 640Zi used in combination with a LeCroy OE425 \cite{teledynelecroy_manual} is provided. The total resolution of the detection system ($\delta T^{\rm FWHM}$) is estimated from the summation in quadrature of each component.  }
    \begin{tabular}{|c|c|c|c|}
      \hline
    Detector Model & $\delta t^{\rm FWHM}$ (ps) & $\delta T^{\rm FWHM}$ (ps) & Measured visibility  \\
          \hline
      LeCroy waveRunner 640Zi \cite{TeledyneLeCroy_WaveRunner6Zi_Datasheet}  & 25 & 34 &$94.4\pm 2.2 \%$ \\
      MPD PDM Series NIM output \cite{MPD_PDM_Datasheet} & 50 & 93 &$92.7\pm 2.7 \%$\\
      MPD PDM Series TTL output& 250 & 262 &$87.0\pm 2.0 \%$\\
      Excelitas SPCM \cite{Excelitas_SPCMAQRH_Datasheet} &350 & 359  & $77.0\pm 2.4 \%$
        \\\hline
    \end{tabular}
    \label{tab:jitter}
\end{table}
\subsection{Moving platforms}
In non-static links, frequency-bin encoded quantum states can experience rapid phase shifts, necessitating a fast compensation system. We describe the time-dependent change in the Heisenberg picture, where an operator $\hat{P}$ is written as
    \begin{align*}
        \hat{P}&=\op{P}{P} && \ket{P}=\sum_i p_{i}e^{i\omega_i t}\ket{\omega_i}.
    \end{align*}
Moreover, for fast moving platforms such as satellites, the Doppler shift must be incorporated. 
For a distance $R(t)$ between the source and receiver, the operator $\hat{P}$ gains a time-dependence
\begin{align}
\hat{P}(t)=\op{P(t)}{P(t)} \quad\quad \ket{P(t)}=\sum_i p_{i}e^{i\omega'_i \frac{R(t)}{t}}\ket{\omega'_i}\approx \sum_i p_{i}e^{i\omega_i(1+\frac{1}{c}\frac{\mathrm{d}R(t)}{\mathrm{d}t})\frac{R(t)}{t}}\ket{\omega_i}
\end{align}
taking both the position-dependent phase shift and the relativistic Doppler effect $\omega'_i=\omega\left(\frac{c+v}{c-v}\right)^{1/2}\approx \omega_i(1+\frac{v}{c})=\omega_i(1+\frac{dR(t)}{cdt})$ into account. For a low-Earth orbit (LEO) satellite, $\frac{dR(t)}{cdt}\approx 2\times 10^{-5}$, which is negligible compared to frequency $\omega_i$.

Moreover, without a precise time reference, the measurement $\tilde{P}(t)$ is a convolution between $\hat{P}(t)$ with a rectangular function $g(t)=1/(2T_r)$ for $t\in[-T_r,T_r]$, depending on the amount of knowledge of $t$ quantified by $T_r$,  
\begin{equation}
\tilde{P}(t)=\int \hat{P}(t-\tau)g(t)d\tau=\frac{1}{2T_r}\int_{-T_r}^{T_r}\hat{P}(t)dt\xRightarrow{T_r\gg\infty}\Delta(P(t))
\end{equation}
where $\Delta(P(t))=\sum_i |p_i|^2\op{\omega_i}{\omega_i}$ is the completely dephasing map.\par 
In a non-static platform, 
the explicit time-dependence of a qubit measurement is $\ket{P(t)}=\ket{0}+e^{i\Delta\omega(1+\frac{dR(t)}{cdt})\frac{R(t)}{c}}\ket{1}$, where the projector evolves due to the transmitter's motion with a phase change $\Delta\omega(1+\frac{dR(t)}{cdt})\frac{R(t)}{c}$, and a phase change rate given by: $\delta f \approx \frac{d}{dt}\left(\Delta\omega(1+\frac{dR(t)}{cdt})\frac{R(t)}{c}\right)$. For a LEO satellite at perigee and approximating its distance to first order as $R(t)=vt$, where $v=\frac{dR(t)}{dt}\approx 6\text{km/s}$ \cite{Liao2017}, a frequency-bin state of $\Delta \omega=2\pi\cdot 260~\text{MHz}$ will experience a phase change rate $\delta f\approx \frac{1}{2\pi} \Delta\omega\frac{v}{c}\approx 5 \text{KHz}$. 
The rapid phase shift is unique to frequency-bins and necessitates continuous phase control to maintain coherence, but can be compensated to better than $99.9\%$ fidelity using GPS time synchronization. A comparison to other degrees of freedom is shown in Table~\ref{table:comparison}. Combined with their on-chip compatibility \cite{clementi_programmable_2023,kues_-chip_2017,borghi2023reconfigurablesiliconphotonicschip}, low SWaP (Size, Weight, and Power) footprint and vibration tolerance frequency-bin quantum sources form an attractive option for satellite-based quantum communication.  
\begin{table}[t]
\centering
\caption{Comparing satellite-based quantum protocols using different degrees of freedom (DOF), sensitivity quantifies how fast the compensation scheme should be.}
\begin{tabular}{|c|c|c|}
\hline
DOF & Sensitivity & Compensation \\
\hline
Polarization\cite{Liao2017}& $<1$ HZ & Polarization Reference + Control \\
Time-bin\cite{Vallone2016} & $<1$ HZ & Phase Modulation \\
Frequency-bin & $\sim5$ KHz & GPS+ Fast Phase Modulation \\
\hline
\end{tabular}
\label{table:comparison}
\end{table}
\section{Experimental demonstration}\label{results}
\begin{figure}
    \centering
    \includegraphics[width=\linewidth]{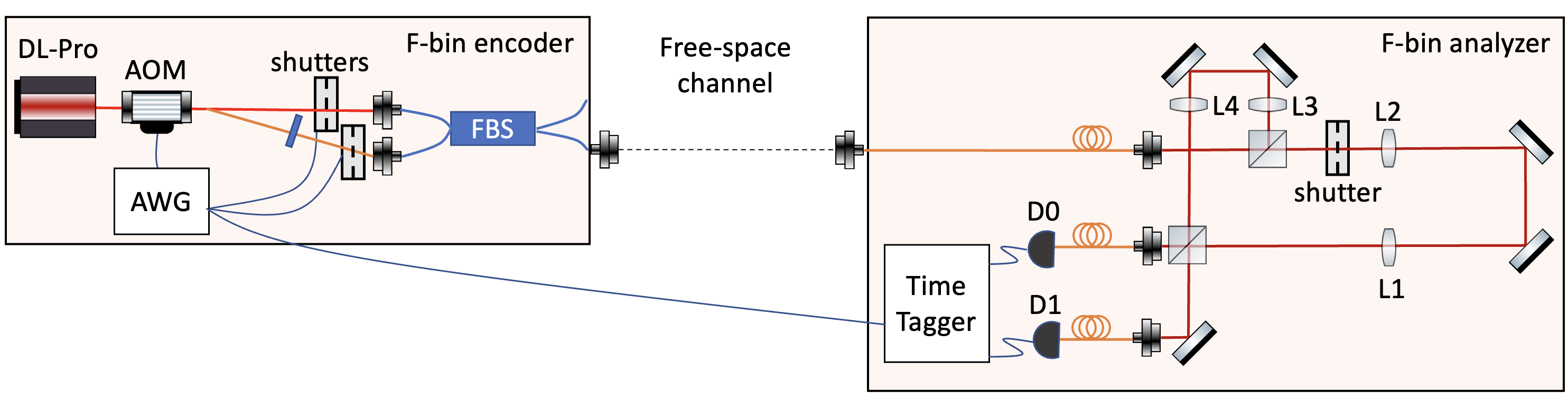}
    \caption{ Experimental setup: The DL-Pro generates 780 nm light which is modulated into two frequency-bins via an acoustic-optical modulator (AOM) driven by an arbitrary waveform generator at 260 MHz. Two mechanical shutters are used to encode the frequency-bins which are spatially combined at a polarization-maintaining fiber beam-splitter and sent across a 2 meter free space bridge to the receiver. The frequency-bin analyzer consists of a field-widened multi-mode interferometer which is used to demultiplex the frequency-bins. The basis choice at the receiver is performed via a mechanical shutter in the long arm of the interferometer. The outputs from the frequency-bin analyzer are coupled into multi-mode fibers connected to Si-APDs. A 500 kHz marker pulse from the AWG provides a time reference to the time-tagging unit.}
    \label{fig:setup}
\end{figure}

We implemented a proof-of-principle demonstration of frequency-bin quantum communication using the setup  depicted in Fig.~\ref{fig:setup}. A continuous-wave laser produces a $ 780$ nm signal, where a portion is detuned by $\Delta\omega$ to a second frequency by an acoustic-optical-modulator (AOM) driven by an RF signal at a frequency of $260$ MHz. Two mechanical shutters encode the frequency-bins. To generate a balanced superposition state ($A_0\simeq A_1$), we recombine both frequency-bins at a fiber beam-splitter with appropriately adjusted coupling efficiencies at each fiber port. The encoded signal is attenuated to single photon level and sent over a two meter free-space channel to a multi-mode frequency-bin analyzer \cite{PhysRevA.97.043847}. Contrary to typical frequency-bin analyzers, our analyzer necessitates no mode filtering nor adaptive optics thus significantly improving the collection efficiency for free-space quantum channels. In the rectilinear basis, the Mach-Zehnder inteferometer operates as a frequency demultiplexer where the channel spacing is determined by the path length difference according to Eq.~\ref{eq:MZI}. The outputs are coupled into multi-mode fibers and detected with Si-APDs. For the superposition basis, a shutter blocks the interferometer's long arm and a beat note is measured at each photon detector. To track the oscillation of the beat note, a $500 \unit{\kilo\hertz}$ marker pulse is sent from the encoder's arbitrary waveform generator (AWG) to the receiver's time-tagger.  
\par
\begin{figure}
    \centering
       \includegraphics[width=\textwidth]{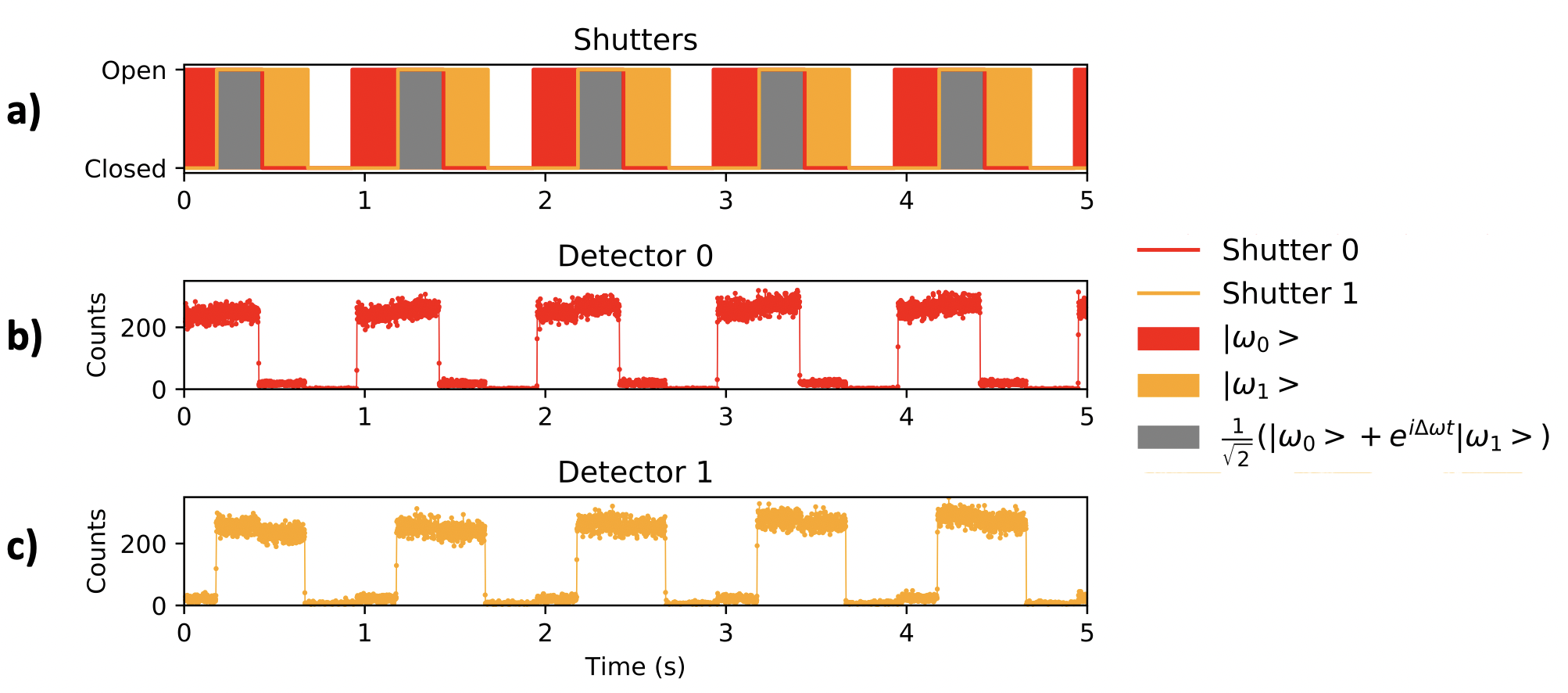}  
   \caption{(a) Frequency-bin encoding scheme. A repeated sequence $\{\ket{\omega_0},\frac{1}{\sqrt{2}}[\ket{\omega_0}+e^{i\Delta\omega t}\ket{\omega_1}], \ket{\omega_1},\ket{vac}\}$ is encoded via two mechanical shutters driven by squares waves $\pi/2$ out of phase. Frequency-bin decoding scheme for the rectilinear basis: $\ket{\omega_0}$ and $\ket{\omega_1}$ are measured on detectors 0 and 1 in (b) and (c) respectively. }
   \label{fig:res}
\end{figure}

\begin{figure}
    \centering
        \begin{subfigure}{.34\textwidth}
      \centering
       \includegraphics[width=\textwidth]{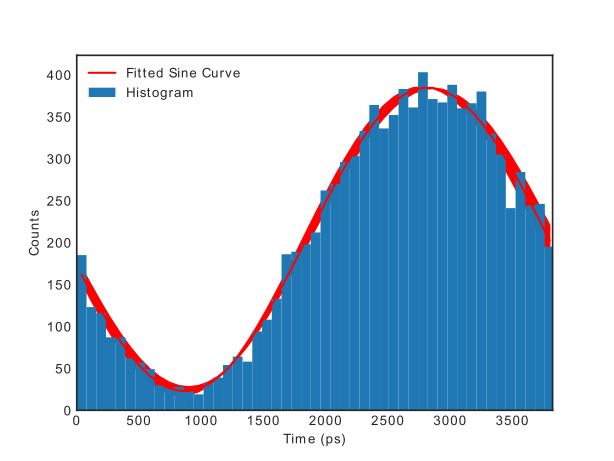}
      \caption{\(X\)-basis encoding 
      }
         \label{fig:nobeat}
          \end{subfigure}
    \begin{subfigure}{.37\textwidth}
      \centering
      \includegraphics[width=\textwidth]{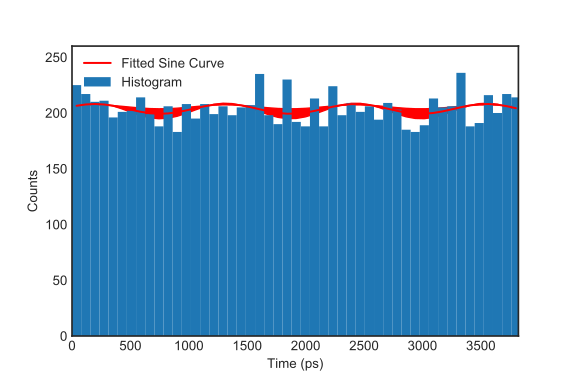}
      \caption{\(Z\)-basis encoding 
      }
         \label{fig:note}
             \end{subfigure}
    \begin{subfigure}{.27\textwidth}
        \includegraphics[width=\textwidth]{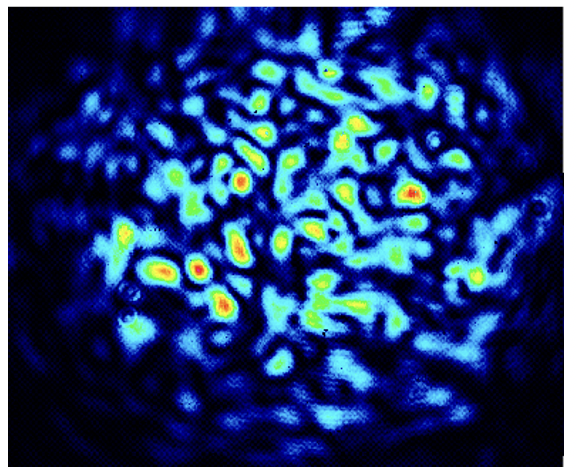}
        \caption{Beam profile}
        \label{fig:beam}
    \end{subfigure}

   \caption{ Frequency-bin decoding scheme for the superposition basis. To evaluate the integrity of the quantum channel, the receiver blocks the long arm of the interferometer and measures a beat note at the photon detector. (a) If the sender prepared a state in the X basis, a beat note corresponding to $\frac{1}{\sqrt{2}}[\ket{\omega_0}+e^{i\Delta\omega t}\ket{\omega_1}]$ can be observed. (b) Whenever the sender encoded in the rectilinear basis only one frequency is received and no beat note is observed. The histogram bin size corresponds to the time-tagger resolution. (c) Image, captured with a
beam-profiling camera (WinCamD-UCD12), of the multi-mode beam in the analyzer, after propagation through 2 meters of free space and a 1 meter long step-index multi-mode
fiber (Thorlabs M43L01).} 
   \label{fig:beat}
\end{figure}

 We demonstrate the feasibility of our scheme by encoding a repeated sequence of $\{0, + ,1 , \mathrm{vac}\}$ depicted in Fig.~\ref{fig:res}a via two mechanical shutters driven at 1 Hz by square waves $\pi/2$ out of phase. The MZI demultiplexes $\ket{\omega_0}$ and $\ket{\omega_1}$ onto detector channels 0 and 1 respectively as can be seen in Fig.~\ref{fig:res}b-c. 
Despite strong spatial distortions (see Fig.~\ref{fig:beam}), we measured a visibility of $85.5\%$ in the \(Z\) basis. While perfect visibility is theoretically possible, the performance is limited is due to imperfect overlap between the interfering beams in the MZI which leads to crosstalk between the inteferometer's output ports.
This leakage is visible in Fig.~\ref{fig:res} in the slight difference with the vacuum floor for detectors 0 (1) when only $\ket{\omega_1}$  ($\ket{\omega_0}$) is transmitted and was used to calculate the visibility according to $V_{\omega_i}=(N_{peak}-N_{leak})/(N_{peak}+N_{leak})$. 
The visibility could be further enhanced through more precise interferometric alignment and improved thermal stability \cite{PhysRevA.97.043847}. For the superposition basis, the beat note shown in Fig.~\ref{fig:nobeat} was measured on a PDM series single photon avalanche photodiode (APD) paired with a UQD Logic16 time-tagger.
If the sender and receiver bases are mismatched, no beat note is observed as is the case in Fig.~\ref{fig:note}. 
The $\sim93$ ps timing resolution of the detection system limited the \(X\)-basis visibility, in Fig.~\ref{fig:res}, to $92.4 \pm 2.7\%$. The experimental parameters for our proof-of-concept demonstration are summarized in Table~\ref{tab:exp}. The encoding rate could be easily improved up to $50$ GHz using off the shelf electro-optic intensity modulators or all-optical switches \cite{ixblue2022mxln,thorlabs2024modulators}. 
While the method outlined in this paper leveraged a prepare-and-measure scheme, it is also compatible with entanglement distribution \cite{tagliavacche2024frequencybinentanglementbasedquantumkey}. In fact, the resource requirement for the latter is reduced as spectrally resolved coincidence counting removes the requirement for the synchronization channel. Significant advancements have been achieved in generating two-photon frequency-bin entanglement.
Techniques such as operating an optical parametric oscillator below threshold \cite{PhysRevLett.91.163602}, filtering broadband parametric down-conversion \cite{xie_harnessing_2015}, and
more recently, the on-chip generation of quantum frequency combs using microring resonators \cite{doi:10.1126/science.aad8532,kues_-chip_2017,Jaramillo-Villegas:17,clementi_programmable_2023} have been explored. Yet these methods have typically been implemented with bin spacing of tens of gigahertz and beyond due to the resolution of dense wavelength division multiplexing filters and wavelength-selective switches \cite{Lu:23}. Using our approach, the resolvable frequency spacing is constrained solely by the timing resolution of the detection system, enabling dense spectral multiplexing for scalable quantum networks \cite{Lingaraju:22,vinet2024reconfigurableentanglementdistributionnetwork}.
\begin{table}[ht]
    \centering
      \captionsetup{justification=centering}
        \caption{Experimental parameters for the proof-of-concept demonstration }
    \begin{tabular}{|c|c|}
    \hline
    $\Delta\omega$ & $1.634\times 10^9$ rad/s\\
    Timing pulse repetition rate & 500 kHz\\
    Shutter repetition rate & 1 Hz\\
     $\omega_0$ visibility & $88.9\pm 2.8\%$\\
     $\omega_1$ visibility & $82.1\pm 3.4\%$\\
              \(Z\)-basis visibility  & $85.5\pm2.2\% $\\
        \(X\)-basis visibility   & $92.4\pm 2.7\% $\\  
  
        \hline
    \end{tabular}

    \label{tab:exp}
\end{table}

\section{Conclusion}\label{conclusion}
Frequency-bin encoding is a promising  approach to quantum networking. Using a purely passive frequency-bin analyzer, we demonstrate the transmission and measurement of frequency-bin encoded photons over a multi-modal free space channel, observing quantum beats without the need for single-mode coupling. Our approach is passive, requiring no spatial filtering or adaptive optics, while maintaining a high optical throughout. The simple receiver design is well-suited for quantum communication over fluctuating channels to moving platforms, including space-based quantum communication. The compatibility of the frequency-bin encoding with solid state devices also enables long-distance quantum information processing. Our multimode receiver allows for free-space links to be incorporated into frequency-bin quantum interconnects, and is thus an enabling step towards global frequency-bin quantum networking.

\begin{backmatter}
\bmsection{Funding}
This work is supported by a joint program of the Natural Sciences and Engineering Research Council of Canada (NSERC) and the European Commission under Grant no. HyperSpace 101070168.  The authors further acknowlege support from the Natural Sciences and Engineering Research Council of Canada (NSERC), the Canadian Foundation for Innovation (CFI), the Ontario Research Fund (ORF) and the Institute for Quantum Computing. SV would like to thank the NSERC PGS-D for personal funding. 
\bmsection{Acknowledgments}
The authors thank Alan Jamison for kindly lending us the acousto-optic modulator used in this experiment, as well as Fabian Steinlechner and Daniele Bajoni for helpful discussions. 
\bmsection{Disclosures}
The authors declare no conflicts of interest related to this article.
\bmsection{Data availability}
The data that support the findings of this study are available upon request to the authors.
\end{backmatter}

\appendix

\bibliography{sample}

\end{document}